\pdfoutput=1 
\documentclass[twocolumn,showpacs,twoside]{revtex4}
\usepackage{amssymb, amsbsy, amsmath, latexsym, dsfont, array, layout, graphicx, mathrsfs}
\usepackage{color}

\newcommand{\ket}[1]{\left|{#1}\right\rangle}
\newcommand{\bra}[1]{\left\langle{#1}\right|}

\begin{document}

\title{Resonator-assisted entangling gate for singlet-triplet spin qubits in nanowire double quantum dots}
\author{Peng Xue$^{1,2}$}
\author{Yun-Feng Xiao$^{3}$}
\affiliation{$^{1}$Department of Physics, Southeast University,
Nanjing 211189, P. R. China}
\affiliation{$^{2}$Institute for
Quantum Information Science, University of Calgary, Alberta T2N 1N4,
Canada}
\affiliation{$^{3}$State Key Lab for Mesoscopic physics,
Peking University, Beijing 100871, P. R. China}
\date{\today}

\begin{abstract}
We propose a resonator-assisted entangling gate for spin qubits with
high fidelity. Each spin qubit corresponds to two electrons in a
nanowire double quantum dot, with the singlet and one of the
triplets as the logical qubit states. The gate is effected by
virtual charge dipole transitions. We include noise in our model to
show feasibility of the scheme under current experimental
conditions.
\end{abstract}


\maketitle

Quantum computing enables some computational problems to be solved
faster than would ever be possible with a classical
computer~\cite{Gro97} and exponentially speeds up solutions to other
problems over the best known classical algorithms~\cite{Sho94}. Of
the promising technologies for quantum computing, solid-state
implementations such as spin qubits in quantum dots
(QDs)~\cite{LD97} and bulk silicon~\cite{Kan98} and charge qubits in
bulk silicon~\cite{ABW+07} and in superconducting Josephson
junctions~\cite{WSB+05} are especially attractive because of
stability and expected scalability of solid-state systems. One of
the most promising qubits in QDs corresponds to a pair of electrons
in a pair of closely-spaced quantum
dots~\cite{Petta05,Johnson05,Taylor07} such that the logical qubit
state $\ket{0}$ is the spin singlet and $\ket{1}$ is one of the spin
triplets. Universal quantum computing is possible if general
single-qubit gates and one entangling two-qubit gate can be
performed.

Here we develop a two-qubit gate for semiconductor quantum
computation based on two-spin states in double quantum dots (DQDs)
realized in the nanowires (NWs) coupled to a superconducting
stripline resonator (SSR). Compared to the previous proposals which
make use of single dots or DQDs defined by a two-dimensional
electron gas (2DEG)~\cite{Ima99,Ima06,Taylor06,Guo09}, our proposal
is more realistic for implementation. It would be difficult to
realize a DQD in a planar resonator with lateral dots, shaped in a
2DEG by surface gates. Because it is difficult to prevent absorption
of microwaves in the 2DEG unless one can make the electric field
non-zero only in the DQD region, which is not realistic
experimentally yet. A more realistic implementation can be done with
NWs as it shows in this work. Very recently, a spin dynamics in InAs
NW QDs coupled to a transmission line via the spin-orbit interaction
has been proposed~\cite{Trif08}. In this work, we propose another
mechanism to achieve an entangling gate between spins inside a SSR,
namely via resonator-assisted interaction which leads to an
efficient coupling between the resonator photon and the effective
electric dipole of DQD and thus eventually entangles the singlet and
one of the triplet spin states.

\begin{figure}[tbp]
   \includegraphics[width=8.5cm]{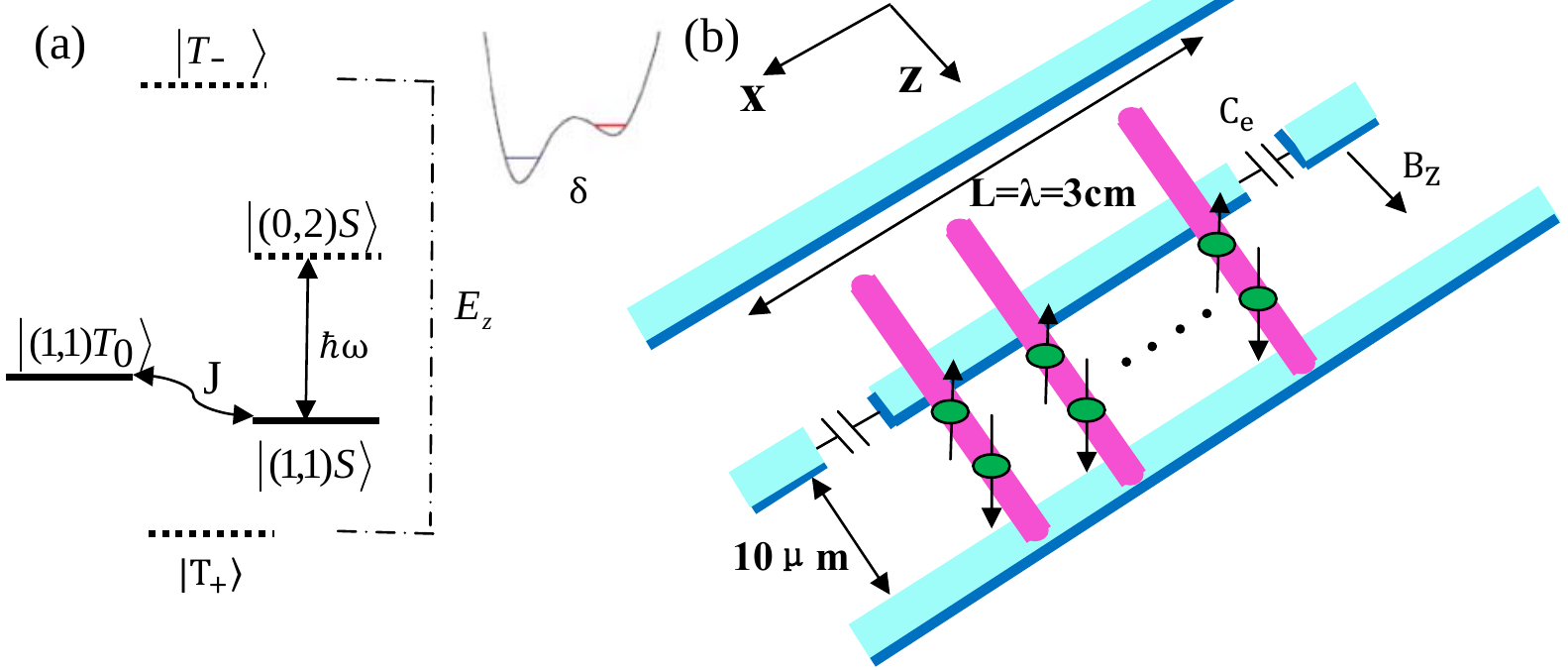}
   \caption{(a) Energy level diagram showing the $(0,2)$ and $(1,1)$ singlets, the three $(1,1)$ triplets
   and the qubit states $\ket{(1,1)S}$ and $\ket{(1,1)T_0}$ with the energy gap $J$ (the exchange energy). Schematic of the double-well
   potential with an energy offset $\delta$ provided by the external electric field. (b) Schematic of
   NW DQDs capacitively coupled to the SSR. The coupling can be switched on and off via the external electric field.
   The DQD confinement can be achieved by barrier materials or by external gates (not shown).}
   \label{fig:energylevel}
\end{figure}

We consider the system with two electrons located in adjacent QDs
inside a semiconductor NW, coupled via tunneling shown in
Fig.~\ref{fig:energylevel}. One of the dots is capacitively coupled
to a SSR. With an external magnetic field $B_z$ along the $z$ axis,
the spin aligned states $\ket{T_+}=\ket{\uparrow\uparrow}$ and
$\ket{T_-}=\ket{\downarrow\downarrow}$, and the spin-anti-aligned
states
$\ket{(1,1)T_0}=(\ket{\downarrow\uparrow}+\ket{\uparrow\downarrow})/\sqrt{2}$
and
$\ket{(1,1)S}=(\ket{\downarrow\uparrow}-\ket{\uparrow\downarrow})/\sqrt{2}$
have energy gaps due to the Zeeman splitting shown in
Fig.~\ref{fig:energylevel}(a). The notation $(n_L,n_R)$ labels the
number of electrons in the left and right QDs of a pair. Considering
a three-level system, we choose the singlet state and one of the
triplet states as our qubit states:
\begin{equation}
\ket{0}\equiv\ket{(1,1)S};\text{ }\ket{1}\equiv\ket{(1,1)T_0},
\label{eq:qubits} \tag{1a}
\end{equation} and the doubly occupied state as an ancillary state
\begin{equation}
\ket{a}\equiv \ket{(0,2)S}. \tag{1b}
\end{equation}

The DQD can be described by an extended Hubbard Hamiltonian
$\hat{H}=(E_\text{os}+\mu)\sum_{i,\sigma}\hat{n}_{i,\sigma}-T\sum_{\sigma}(\hat{a}^\dagger_{L,\sigma}\hat{a}_{R,\sigma}+\text{hc})
+U\sum_i\hat{n}_{i,\uparrow}\hat{n}_{i,\downarrow}
+W\sum_{\sigma,\sigma'}\hat{n}_{L,\sigma}\hat{n}_{R,\sigma'}+\delta\sum_{\sigma}(\hat{n}_{L,\sigma}-\hat{n}_{R,\sigma})$
for $\hat{a}_{i,\sigma}$ ($\hat{a}^\dagger_{i,\sigma}$) annihilating
(creating) an electron in a QD~$i\in\{L,R\}$ with
spin~$\sigma\in\{\uparrow,\downarrow\}$,
$\hat{n}_{i,\sigma}=\hat{a}^\dagger_{i,\sigma}\hat{a}_{i,\sigma}$ a
number operator, and $\delta$ an energy offset yielded by the
external electric field. The first term corresponds to on-site
energy~$E_\text{os}$ plus site-dependent field-induced
corrections~$\mu$. The second term accounts for $i \leftrightarrow
j$ electron tunneling with rate~$T$. The third term is the on-site
charging cost~$U$ to put two electrons with opposite spin in the
same dot, and the fourth term corresponds to inter-site Coulomb
repulsion. In the basis $\{\ket{0},\ket{a}\}$, the Hamiltonian can
be reduced as
\begin{equation}
\hat{H}_\text{d}=-\delta\ket{a}\bra{a}+T\ket{0}\bra{a}+\text{hc}.
\label{eq:deduceHam1} \tag{2}
\end{equation}
With the energy offset $\delta$, degenerate perturbation theory in
the tunneling $T$ reveals an avoided crossing at this balanced point
which occurs at the left-most avoided crossing between $\ket{0}$ and
$\ket{a}$ with an energy gap $\omega=\sqrt{\delta^2+4T^2}$.

The essential idea is to use an effective electric dipole moment
associated with singlet states $\ket{0}$ and $\ket{a}$ of a NW DQD
coupled to the oscillating voltage associated with a SSR shown in
Fig.~\ref{fig:energylevel}(b). Whereas the qubit state $\ket{1}$ is
decoupled to the SSR due to the large energy gap $J$. We consider a
SSR with the length $L$, the capacitance per unit length $C_0$ and
the characteristic impedance $Z_0$. A capacitive coupling $C_c$
between the NW DQD and SSR causes the electron charge state to
interact with excitations in the transmission line. We assume that
the dot is much smaller than the wavelength of the resonator
excitation, so the interaction strength can be derived from the
electrostatic potential energy of the system
$\hat{H}_\text{int}=e\hat{V}v\ket{(0,2)S}\bra{(0,2)S}$ with $e$ the
electron charge,
$\hat{V}=\sum_{n}\sqrt{\frac{\hbar\omega_n}{LC_0}}(\hat{c}_n+\hat{c}_n^\dagger)$
the voltage on the SSR near the left dots, $\hat{c}_n$
($\hat{c}^\dagger_n$) the creation (annihilation) operator for the
SSR mode $k_n=[(n+1)\pi]/L$, $v=C_c/C_\text{tot}$, and
$C_\text{tot}$ the total capacitance of the DQD. The fundamental
mode frequency of the SSR is $\omega_0=\pi/LZ_0C_0\approx \omega$.
The SSR is coupled to a capacitor $C_e$ for writing and reading the
signals. Neglecting the higher modes and working in the rotating
frame with the rotating wave approximation, we obtain an effective
interaction Hamiltonian as
\begin{equation}
\hat{H}_\text{int}=g(\hat{c}\ket{a}\bra{0}+\text{hc})
\label{eq:intHam} \tag{3}
\end{equation}
with the effective coupling coefficient
\begin{equation}
g=\frac{1}{2}ev\frac{1}{LC_0}\sqrt{\frac{\pi}{Z_0\hbar}}\sin2\vartheta
\label{eq:g} \tag{4}
\end{equation}
with $\vartheta=\frac{1}{2}\tan^{-1}(\frac{2T}{\delta})$.

The interaction between the SSR and qubit states is switchable via
tuning the electric field. In the case of the energy offset yielded
by the electric field $\delta\approx 0$, we obtain the maximum value
of the coupling between the SSR and qubit states. Whereas $\delta\gg
T$, $\vartheta$ tends to $0$, the interaction is switched off.

We consider there are two NW DQDs coupled to the SSR. If both of the
DQDs are in the state $\ket{1}$, the incoming pulses which are
resonant with the bare resonator mode and performed in the limit
with $\tau \gg 1/\kappa$ (here $\tau$ is the pulse duration and
$\kappa$ is the resonator decay rate) are resonantly reflected by
the bare resonator mode with a flipped global phase of $\pi$ from
the standard quantum optics calculation~\cite{Wall94,atom}. For the
other three cases including that the DQDs are in the states
$\ket{00}$, $\ket{01}$ and $\ket{10}$ the frequency of the dressed
resonator mode is significantly detuned from the frequency of the
incoming pulses and the frequency shift has a magnitude comparable
with $g$. Thus the SSR functions as a mirror and the shape and
global phase of the reflected pulses remain unchanged.

We now go to a detailed description of the resonator-assisted
interaction with the input field in an even coherent state
$\ket{\alpha}_- =N_-(\ket{\alpha}-\ket{-\alpha})$, where $N_-$ is
normalization constant and $\ket{\alpha}$ is a coherent state.
Recently, this novel state of light has been generated and
characterized by a non-positive Wigner function experimentally
\cite{Nielsen06}. The SSR output $\hat{c}_\text{out}$ is connected
with the resonator mode $\hat{c}$ via the following relations
\begin{equation}
\dot{\hat{c}}(t)=-i\left[\hat{c}(t),\hat{H}\right]-(i\Delta+\frac{\kappa}{2})\hat{c}(t)-\sqrt{\kappa}\hat{c}_\text{in}(t);
\label{eq:ac} \tag{5}
\end{equation}
\begin{equation}
\hat{c}_\text{out}(t)=\hat{c}_\text{in}(t)+\sqrt{\kappa}\hat{c}(t).
\label{eq:acout}\tag{6}
\end{equation}
Here $\Delta$ denotes the detuning of the resonator field mode
$\hat{c}(t)$ from the input pulse $\hat{c}_\text{in}(t)$ with the
standard communication relation
$\big[\hat{c}_\text{in}(t),\hat{c}_\text{in}(t')\big]=\delta(t-t')$.
We get the following results. If both DQDs are in the state
$\left\vert {1}\right\rangle $, the Hamiltonian $\hat{H}$ is not
active. Based on Eqs.~(\ref{eq:ac}) and (\ref{eq:acout}) we obtain
$\hat{c}_{\text{out}}^{11}(t)=-\hat{c}_{\text{in}}^{11}(t)$ if the
resonant interaction satisfies $\Delta=0$ and the input pulse shape
changes slowly with time $t$ compared with the resonator decay rate
$\kappa$. That means if the state of both DQDs is in $\left\vert
{11}\right\rangle$, the output field acquires a phase $\pi$ after
the interaction. In other input cases, however, there are effective
detunings of the dressed resonator mode from the input pulse
$\pm\sqrt{2}g$ for $\left\vert {00}\right\rangle $ and $\pm g$ for
$\left\vert {01}\right\rangle $ ($\left\vert {10}\right\rangle $)
respectively and the input-output equations
$\hat{c}_\text{out}^{mn}(t)=\xi_{mn}\hat{c}_\text{in}^{mn}(t)$
($m,n=0,1$) where $\xi_{11}=-1$, $\xi_{00}=(8s-1)/(8s+1)$ and
$\xi_{01}=\xi_{10}=(4s-1)/(4s+1)$ with $s=g^{2}T_1/\kappa$ ($T_1$ is
charge relaxation time). If $s\gg1$ is satisfied, $\xi_{00}$,
$\xi_{01}$ and $\xi_{10}$ tend to $1$. Thus for an arbitrary initial
state, after the resonator-assisted interaction and tracing out the
resonator field, we implement a controlled phase flip (CPF) gate
$e^{i\pi\ket{11}\bra{11}}$ on the spin states of the two DQDs and
the gate time is about $t_{\text{cpf}}\sim\tau$ ($\tau$ is the pulse
duration).

\begin{figure}[tbp]
   \includegraphics[width=8.5cm]{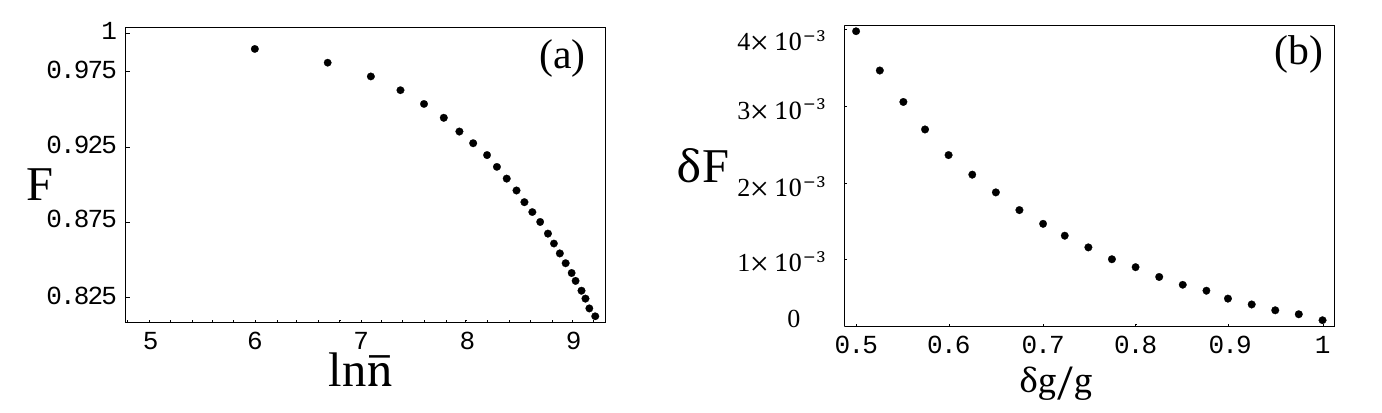}
   \caption{(a)The fidelity of the CPF gate versus
   the mean photon number of the input pulse with the pulse
   duration $\tau=10/\kappa$, and (b) it changes with $\delta g/g$. Here we
   choose the realistic parameters $(g,\kappa,1/T_1)/2\pi=(120,100,1)$MHz.}
   \label{fig:fidelity}
\end{figure}

Now we analyze the feasibility of the proposal with the elongated
QDs oriented along the NW. QDs have been realized within NWs in
various material systems~\cite{nanowires}. A realization of DQDs
defined using local gates to electrostatically deplete InAs NWs
grown by chemical beam epitaxy was reported~\cite{Fasth05}. The
quantum-mechanical tunneling $T$ between the two QDs is about
$0-150\mu$eV~\cite{Fasth05}. Thus at the optimal point
$\delta\approx 0$ where the coupling is strongest, the energy gap
between the singlets is about $\omega\sim2T\simeq 0-72$GHz. The SSR
can be fabricated with existing lithography
techniques~\cite{Wallraff04}. The dots can be placed within the SSR
formed by the transmission line to strongly suppress the spontaneous
emission. A small-diameter ($d\sim65$nm), long-length ($l\sim270$nm)
and $g^*=-13$~\cite{MT05} InAs NW is positioned perpendicularly to
the transmission line and containing DQDs that are elongated along
the NW shown in Fig.~1(b). To prevent a current flow, the NW and the
transmission line need to be separated by some insulating coating
material obtained for example by atomic layer deposition. We assume
that the SSR is $3$cm long and $10\mu$m wide, $Z_0=50\Omega$, which
implies for the fundamental mode $\omega_0=2\pi\times10$GHz. The
external magnetic field along the $z$ axis is about $B_z=1$T to make
sure the energy splitting $E_z=g^*\mu_B B_z$ between the two triplet
states $\ket{T_{\pm}}$ is larger than $\hbar\omega$. In practice,
the careful fabrication permits a strong coupling capacitance, with
$v\approx 0.2 $~\cite{Fasth05}, so that the coupling coefficient
$g\sim 2\pi\times120$MHz is achievable due to the numerical
estimations in Eq.~(\ref{eq:g}). The frequency $\omega_0$ and
coupling coefficient $g$ can be tuned via $LC_0$. In order to
implement gates on a fast time scale, the system works in a weak
coupling regime. In the bad cavity limit, we have $g\sim\kappa$,
where $\kappa=\omega_0/Q\approx 2\pi\times 100$MHz with the quality
factor of the SSR $Q=100$~\cite{Frunzio05}. Considering the effect
of photon assisted tunneling in our system, which is harmful because
it destroys the qubit by lifting spin-blockade. To avoid this, one
needs to close enough the tunneling barriers to the leads.

We address now the issue of relaxation and decoherence of our
system. There are three types of contributions to the relaxation
processes, one arising from the finite decay rate of the SSR, one
from the intrinsic decoherence of the spin states, and the other one
from the charge-based dephasing and relaxation which occur during
the gate operation involving the electric dipole between $\ket{a}$
and $\ket{0}$.

For the charge relaxation time $T_1$, the decay is caused by
coupling qubits to a phonon bath. With the spin-boson model, the
perturbation theory gives an overall error rate from the relaxation
and incoherent excitation, with which one can estimate the
relaxation time $T_1\sim1\mu$s~\cite{Taylor06}. That is studied in
great details for the GaAs QDs in 2DEG and a similar rate is
expected for NW QDs.

The charge dephasing $T_2$ rises from variations of the energy
offset $\delta(t)=\delta+\varepsilon(t)$ with
$\langle\varepsilon(t)\varepsilon(t')\rangle=\int d\omega
S(\omega)e^{\text{i}\omega(t-t')}$, which is caused by the low
frequency fluctuations of the electric field. The gate bias of the
qubit drifts randomly when an electron tunnels between the metallic
electrode. At the zero derivative point, compared to a bare
dephasing time $T_b=1/\sqrt{\int d\omega S(\omega)}$, the charge
dephasing is $T_2\sim \omega T_b^2$ near the optimal point
$\delta=0$. The bare dephasing time $T_b\sim 1$ns was observed
in~\cite{Hayashi04}. Then the charge dephasing is about
$T_2\sim10-100$ns. Using quantum control techniques, such as better
high- and low-frequency filtering of electronic noise, $T_b$
exceeding $1\mu$s was observed in 2DEG~\cite{Petta05} (we assume a
similar result for the present case), which suppresses the charge
dephasing.

The hyperfine interactions with the host nuclei cause nuclear
spin-related dephasing $T_2^*$. The hyperfine field can be treated
as a static quantity, because the evolution of the random hyperfine
field is several orders slower than the electron spin dephasing. In
the operating point, the most major decoherence due to the hyperfine
field is the dephasing between the singlet state $\ket{0}$ and one
of the triplet states $\ket{1}$. By suppressing nuclear spin
fluctuations, the spin dephasing time can be obtained by
quasi-static approximation as $T_2^*=1/g\mu_B\langle\Delta
B_n^z\rangle_\text{rms}$, where $\Delta B_n^z$ is the nuclear
hyperfine gradient field between two coupled dots and rms means a
root-mean-square time-ensemble average. A measurement of the spin
dephasing time $T_2^*\sim4$ns was demonstrated in~\cite{Trif08} and
we expect that coherently driving the qubit will prolong the $T_2^*$
time up to $1\mu$s and with echo up to $10\mu$s~\cite{Petta05}.

The quality factor $Q$ of the SSR in the microwave domain can be
achieved up to $10^6$~\cite{Wallraff04}. In practice, the local
external magnetic field $B_z=1$T reduces the limit of
Q~\cite{Frunzio05}. However, in our proposal the low Q SSR is good
for implementing an entangling gate with short operation time. The
dissipation of the SSR $\kappa$ leads the decay time about $10$ns,
which will cause photon loss. Due to the photon loss, the amplitude
of the output field $\alpha_{mn}$ for $m,n=0,1$ representing the
different initial states of the DQDs is probably different from the
input amplitude $\alpha$ and usually $|\alpha_{mn}|<|\alpha|$. The
fundamental source of photon loss in the SSR can be qualified by the
parameter
$\eta=1-\text{min}\{|\alpha_{mn}/\alpha|^2\}\propto\kappa/g^2T_1$.

Except for the photon loss, the function distortion is another
reason which causes the noise. The input to the SSR is a coherent
field which can be described as
$\ket{\alpha}_\text{in}=e^{\alpha\int_0^\tau
f_\text{in}(t)\hat{c}^\dagger_\text{in}(t)\text{d}t}\ket{vac}$,
where $f_\text{in}(t)$ describes the input pulse shape with
normalization $\int_0^{\tau}|f_\text{in}(t)|^2\text{d}t=1$ with the
pulse duration $\tau$. We calculate the output pulse shapes
$f^{mn}_\text{out}(t)$ from the expectation value of the
input-output equation (\ref{eq:acout})
$\alpha_{mn}f^{mn}_\text{out}(t)=\alpha
f_\text{in}(t)+\sqrt{\kappa}\langle\hat{c}(t)\rangle$. The
expectation value of the resonator mode $\hat{c}(t)$ can be obtained
by solving the corresponding master equation
\begin{align}
\dot{\rho}=&-i\big[\hat{H}_\text{eff},\rho\big]+\frac{\kappa}{2}(2\hat{c}\rho\hat{c}^\dagger-\hat{c}^\dagger\hat{c}\rho
-\rho\hat{c}^\dagger\hat{c})\nonumber\\
&+\frac{1}{2T_1}\big(2\hat{\sigma}_-\rho\hat{\sigma}_+-\hat{\sigma}_+\hat{\sigma}_-\rho-\rho\hat{\sigma}_+\hat{\sigma}_-\big)
\label{master}\tag{7}
\end{align}
with $\hat{\sigma}_+=\ket{a}\bra{0}$,
$\hat{\sigma}_-=\ket{0}\bra{a}$ and the effective Hamiltonian
$\hat{H}_\text{eff}=g\hat{\sigma}_+\hat{c}+i\sqrt{\kappa}\langle
\hat{c}_\text{in}(t)\rangle \hat{c}+\text{hc}$. By solving the
master equation we obtain
$\langle\hat{c}(t)\rangle=\text{tr}(\rho\hat{c}(t))$. Then we can
determine the output amplitude $\alpha_{mn}$ and the corresponding
pulse shape $f_\text{out}^{mn}(t)$ and define the mismatching of the
input and output pulses for the initial state $\ket{mn}$ as
$\epsilon_{mn}=1-\int\int
f^*_\text{in}(t')f^{mn}_\text{out}(t)\text{d}t\text{d}t'$.

Now we analyze the fidelity of the CFP gate under the influence of
some practical sources of noise. For the initial state of the system
$\ket{\Phi}_\text{in}$ with the input pulse
$\ket{\phi}_\text{in}\propto \big[e^{\alpha\int_0^\tau
f_\text{in}(t)\hat{c}^\dagger_\text{in}(t)\text{d}t}-e^{-\alpha\int_0^\tau
f_\text{in}(t)\hat{c}^\dagger_\text{in}(t)\text{d}t}\big]\ket{vac}$
the input pulse corresponding to the double-dot state $\ket{mn}$. By
applying a CFP gate, the output state of the system can be written
as $\ket{\Phi}_\text{out}$ with the output pulse
$\ket{\phi}_\text{out}^{mn}$ with a shape $f_\text{out}^{mn}(t)$ and
amplitude $\alpha_{mn}$ different from the input:
$\ket{\phi}_\text{out}^{mn}\propto \big[e^{\alpha_{mn}\int_0^\tau
f_\text{out}^{mn}(t)\hat{c}^\dagger_\text{out}(t)\text{d}t}-e^{-\alpha_{mn}\int_0^\tau
f_\text{out}^{mn}(t)\hat{c}^\dagger_\text{out}(t)\text{d}t}\big]\ket{vac}$.
Then the fidelity can be defined as
\begin{widetext}
\begin{align}
&F=\bra{\Phi_\text{ID}}\rho(t_\text{cpf})\ket{\Phi_\text{ID}}\tag{8}\\
&=\big|\big\{\sum_{mn=00,01,10}e^{-\frac{1}{2}|\alpha|^2
[(1-\epsilon_{mn})^2+(1-\eta_{mn})-2\xi_{mn}\sqrt{1-\eta_{mn}}(1-\epsilon_{mn})]}
+e^{-\frac{1}{2}|\alpha|^2[(1-\epsilon_{11})^2+(1-\eta_{11})+2\xi_{11}\sqrt{1-\eta_{11}}(1-\epsilon_{11})]}\big\}/4\big|^2,\nonumber
\end{align}
\end{widetext}
where the ideal output state is
$\ket{\Phi_\text{ID}}=e^{i\pi\ket{11}\bra{11}}\ket{\Phi}_\text{in}$.

We investigate the fidelity under typical experimental
configurations and it is shown in Fig.~\ref{fig:fidelity} as a
function of the mean photon number of the input state with the
realistic parameters $(g,\kappa,1/T_1)/2\pi=(120,100,1)$MHz. In the
calculation, we have assumed a Gaussian shape for the input pulse
with $f_\text{in}(t)\propto e^{[-(t-\tau/2)^2/(\tau/5)^2]}$ with the
pulse duration $\tau=10/\kappa$. We obtain a high fidelity up to
$0.99$ for these parameters and the coherent input pulse with a
remarkable amplitude $\alpha\sim 20$. Furthermore, the fidelity $F$
is insensitive to the variation of the coupling coefficient $g$
caused by the fluctuations in the DQD positions and the energy gap
$\delta$. The change of the fidelity $\delta F$ is about $10^{-3}$
for $g$ varying to $g/2$. Furthermore, we can estimate the time
scaling for the CPF gate operation $t_\text{cpf}\sim \tau=100$ns
with fidelity $0.99$, which is shorter compared to the decoherence
time.

In summary, we propose a resonator-assisted entangling gate for
singlet-triplet spin qubits in DQDs, which exploits virtual
charge-qubit transitions for double-dot pairs capacitively coupled
to a SSR. Because of the switchable coupling between the double-dot
pairs and the SSR, we can apply this entangling gate on any two
qubits without affecting others, which is not trivial for
implementing scalable quantum computing and generating large
entangled states. The fidelity of this gate is studied including all
kinds of major decoherence, with promising results for reasonably
achievable experimental parameters. The feasibility of this scheme
is characterized through exact numerical simulations that
incorporate various sources of experiment noise and these results
demonstrate the practicality by way of current experimental
technologies.

\textit{Acknowledgements}---We wish to thank Barry C. Sanders,
Michel Pioro-Ladri\`{e}re and Alexandre Blais for valuable
discussions. This work has been supported by NSFC 10944005, NSFC
10821062, NSERC, MITACS, CIFAR, QuantumWorks and iCORE.

\end{document}